\documentstyle[aps,prl]{revtex}
\tighten
\input epsf
\newcommand{\journal}[4]{{{\sl #1}} {\bf #2}, {#3} (#4)}
\newcommand{\mprb}[3]{\journal{Phys.~Rev.~B}{#1}{#2}{#3}}
\newcommand{\mpre}[3]{\journal{Phys.~Rev.~E}{#1}{#2}{  #3}}
\newcommand{\mpra}[3]{\journal{Phys.~Rev.~A}{#1}{#2}{#3}}
\newcommand{\mprl}[3]{\journal{Phys.~Rev.~Lett.~}{#1}{#2}{#3}}

\begin{document}
\twocolumn

\title{Entropy-vanishing transition and glassy dynamics in frustrated spins}

\author{Hui Yin and Bulbul Chakraborty}

\address{Martin Fisher School of Physics, Brandeis University, Waltham, MA
02454.}

\date{\today}

\maketitle

\begin{abstract}
In an effort to understand the glass transition, the dynamics  of a non-randomly
frustrated spin model has been analyzed. The phenomenology of the spin model 
is
similar to that
of a supercooled liquid undergoing the glass transition. 
The slow dynamics can
be associated with the presence of extended string-like structures 
which demarcate regions of fast spin flips.  An entropy-vanishing transition, with the string density as the order parameter, is related to the observed glass transition in the spin
model.
\end{abstract}
\vspace{0.2in}
The glass transition in supercooled liquids is heralded by anomalously slow
relaxations with a time scale diverging as the liquid freezes into 
the glassy state\cite{Glassrev}.  In recent years,
there have been careful
experimental and theoretical studies aimed at
understanding the structural aspects of this transition.  The presence and nature of
dynamical heterogeneities near the glass transition has been a dominant
underlying theme of both simulations\cite{Glotzer,Klein} and experiments\cite{Ediger,Weitz}.  
Simulations in Lennard-Jones
liquids\cite{Glotzer} have shown the existence of string-like dynamical heterogeneities and 
similar structures have been observed directly in colloidal glasses\cite{Weitz}.  
In the Adam-Gibbs scenario, the glass transition is related to a phase
transition accompanied by the vanishing of configurational entropy\cite{Adam,Mezard}.
An explicit connection between (a) dynamical heterogeneities, (b)
anomalous relaxations and (c) the Adam-Gibbs scenario would provide useful 
insight into the nature of the glass transition. In this work, we present our
analysis of a simple model where there are naturally
occurring dynamical heterogeneities in the form of strings 
and where there is an entropy vanishing
transition involving these structures.  Monte Carlo simulations of the model
show that there is a glass-like transition with diverging time scales and
an anomalously broad relaxation spectrum. 
Analysis of the simulation results provides strong evidence that the entropy-vanishing
transition underlies the observed dynamical behavior.
 
\paragraph{Model}One of the simplest  non-randomly frustrated spin models is
the triangular-lattice Ising antiferromagnet (TIAFM).  The TIAFM has an exponentially
large number of ground states and has a zero-temperature critical
point\cite{Henley,wannier,houtappel}. The model studied in this letter is the {\it compressible}
TIAFM (CTIAFM) in which the coupling of the spins to the elastic strain fields removes the exponential degeneracy
of the ground-state.  We solve the CTIAFM exactly within the ground-state
ensemble of the TIAFM and show that there is an entropy-vanishing transition
which involves extended structures. We 
then present results of simulations which indicate that the entropy-vanishing
transition leads to glassy dynamics.

The Hamiltonian of the CTIAFM is:
\begin{eqnarray}
H &=& J{\sum}_{<ij>}S_i S_j -\epsilon J
{\sum}_{\alpha}e_{\alpha}{\sum}_{<ij>_{\alpha}}S_i S_j \nonumber \\ 
&+& N{E \over 2}{\sum}_{\alpha}e_{\alpha}^{2}~.
\label{model}
\end{eqnarray}
Here $J$, the strength of the anti-ferromagnetic coupling, is modulated by 
the presence of the second term which defines a coupling between the spins and
the homogeneous  strain fields ${e_{\alpha}, \alpha = 1,2,3}$, along the three
nearest-neighbor directions on the triangular lattice.
The last term stabilizes the unstrained lattice. The total number of spins in
the system is given by $N$.
The ground-state of the CTIAFM is a three-fold
degenerate striped phase where up and down spins alternate between rows and there is 
a shear distortion characterized by $e_1 = e$ and $e_2 = e_3 = -e$ if the rows
are along the direction of $e_1$\cite{Kardar,Lei1}.  
Within the manifold of the TIAFM ground-states, the competition between energy gained from the lattice
distortions and the extensive entropy of the TIAFM ground-states leads to an entropy-vanishing
transition. The order parameter associated with this entropy-vanishing transition 
is the density of extended string-like structures which characterize the TIAFM
ground states.

\begin{figure}[bthp]
\hspace{0.4in}
\epsfxsize=2.0in \epsfysize=1.8in
\epsfbox{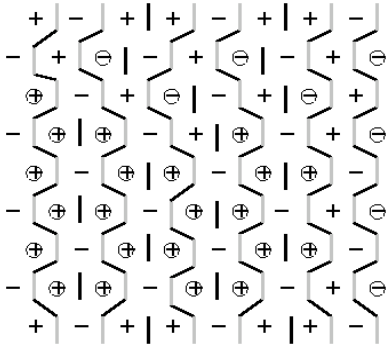}
\vspace{0.2in}
\caption{String representation of a TIAFM ground state. Dark bonds of the dual
honeycomb lattice ({\it cf} text) are the
dimers which divide unsatisfied pairs of spins. Light bonds define the
``standard'' dimer configuration.  The strings are made up of these two
types and extend across the system.  The fast 
spins, the ones 
with $3-3$ coordination, have been encircled.}  
\label{string_high}
\end{figure}

The string picture of the TIAFM ground states derives from a well-known mapping
of these states to dimer coverings\cite{Dhar,Zeng}.  In a ground state, there
is one unsatisfied bond per triangular plaquette and the dimers are the
filled-in  bonds of the dual honeycomb lattice that cross the unsatisfied bonds of the
triangular lattice, as shown in Fig.(\ref{string_high}).  Superposing a dimer 
configuration on a ``standard'' dimer configuration where all the dimers are
vertical\cite{Dhar,Zeng} leads to a string configuration ({\it cf} Fig.(\ref{string_high})). 
Assuming
spin-flip dynamics for the moment, the only spins that can be changed while the
system remains in the ground-state manifold, are the ones which have a
coordination of 3-3 (3 satisfied and 3 unsatisfied bonds).  These are the
fast spins in the system, and,
as
shown in Figure (\ref{string_high}), are located at isolated kinks on the
strings.  The strings, therefore, play the role of dynamical heterogeneities in
this lattice model.

\begin{figure}[tbhp]
\hspace{0.2in}
\epsfxsize=2.0in \epsfysize=2.0in
\epsfbox{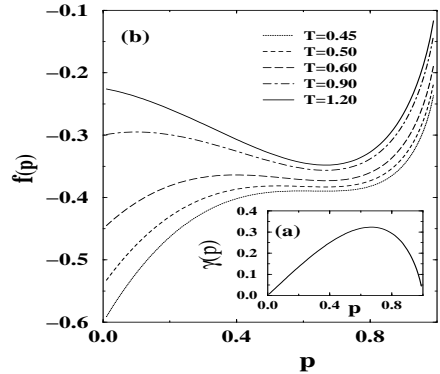}
\caption{(a) Entropy as a function of the string density from the work of Dhar et 
al.(\protect \cite{Dhar}) (b)The dimensionless  free energy $f(p)$ for $\mu =
0.18$. $T^*$ for this value of
$\mu$ is 0.397.  Temperature is measured in units of $1/k_{B}$.}  
\label{gammap}
\end{figure}

\paragraph{Exact results} We can solve the CTIAFM exactly within the restricted 
spin ensemble of the ground states of the TIAFM.  All the states in this ensemble
can be 
classified according to the string
density $p=N_s/L$ where $L$ is the linear dimension of the sample and $N_s$ is the
number of strings\cite{Dhar}.    The
number of spin states ($\Omega(p)$) belonging to a particular string-density
sector $p$ has been
shown to be exponentially large\cite{Dhar}: 
$\Omega(p)=\exp (N\gamma(p))$, with 
$N = L \times L$ being the total number of spins. As shown in
Fig. (\ref{gammap}), 
the entropy $\gamma (p)$ has a peak at $p=2/3$\cite{Dhar}. In the CTIAFM (Eq. \ref{model}), the strain $e_{\alpha}$
couples to 
$\langle{S_i S_j}\rangle_{\alpha}$.  This average counts the number 
of strings along the $\alpha$ direction, {\it i.e.} the number of strings
obtained by taking an overlap with the standard dimer state with all the dimers
perpendicular to the $\alpha$ direction. 
Under periodic boundary conditions, only one of the three string densities is  
independent.

The strain field appears in the
Hamiltonian as a purely Gaussian variable and can be integrated out, and the
CTIAFM 
energy per spin, {\it in the restricted ensemble},  can  be written in terms of
the one, independent 
string density $p$:
\begin{equation}
E(p) = -(\mu /2)((1-2p)^2 +2(1-p)^2))~.
\end{equation}
Here $\mu ={\epsilon}^2 J^2/E$.  The energy
function $E(p)$ distinguishes different string sectors and is minimized by
$p=0$. The entropy of the ground states, $\gamma (p)$, on the other hand favors
the $p =2/3$ sector and the competition between energy and entropy leads to the
possibility of a phase transition.
In the thermodynamic
limit,  the partition function $Z = {\sum}_p \exp^{-N f(p)}$, is dominated by
the string density which minimizes
$f(p)=\beta E(p) -\gamma(p)$, where $\beta$ is the inverse temperature.  The exact free energy corresponds to this
minimum value of $f(p)$ and the only relevant coupling constant in the problem is
$\beta \mu$.
For small values of the coupling constant, 
the function $f(p)$ shown in Fig (\ref{gammap}(b)), has only one minimum at
$p=2/3$.  As the coupling constant is increased, 
this minimum stays pinned at $2/3$ and a second minimum
starts developing at $p=0$. The $p=0$ state stays metastable until at  $\mu/T_1
=(3/4)*\gamma (2/3) \simeq 0.24$ there is a first-order transition from the 
$p=2/3$ state to the $p=0$ state.  The $p=2/3$ state loses its metastability at
a larger coupling given by $\mu/T^* = \sqrt{3}\pi/12$.   At $T^*$, the entropy
vanishes and the order parameter
shows a discontinuous change from $p=2/3$ to $p=0$.  This transition is reminiscent of the
transitions observed in p-spin spin glasses\cite{Mezard,Kirkpatrick}. 

These exact results show that in the CTIAFM, there is an entropy vanishing
transition which defines the limit of stability of the homogeneous, liquid-like
$p = 2/3$ state.  In the Adam-Gibbs 
scenario, such a transition underlies the glass transition. The exact results
are  valid for the CTIAFM acting within the ground-state ensemble of the TIAFM.
Defects\cite{Blote}, 
which correspond to triangles with all three bonds unsatisfied, can take the system
out of the ground-state manifold. For low defect densities, however, it is
possible that the entropy-vanishing transition survives in some form and leads
to slow glassy dynamics.  We have investigated this scenario by performing
Monte Carlo simulations of the CTIAFM.

\paragraph{Simulations} The parameters of the model were
chosen to be 
$J = 1$, $\epsilon = 0.6$ and $E = 2$.  These values yield a small value of
$\mu$ ($=0.18$) and ensures that at the transition, $T^*$ ($=0.397$), the defect density is 
low.  The average defect number-density was measured to be $\simeq
0.04{\%}$ at $T=0.45$. 
We used Monte Carlo simulations to study  the dynamics  of the supercooled
state following instantaneous quenches to
temperatures below $T_1 =0.75$.
Spin-exchange kinetics was extended to include moves
which attempted changes of the 
strain fields $e_{\alpha}$. Details of the
simulation algorithm have been published earlier\cite{Lei1}. System sizes
ranged from 48x48 up to 120x120. Unless otherwise stated, the results presented
in this letter were obtained from 96x96 systems.

\begin{figure}[hbtp]
\epsfxsize=2.5in \epsfysize=3.0in
\epsfbox{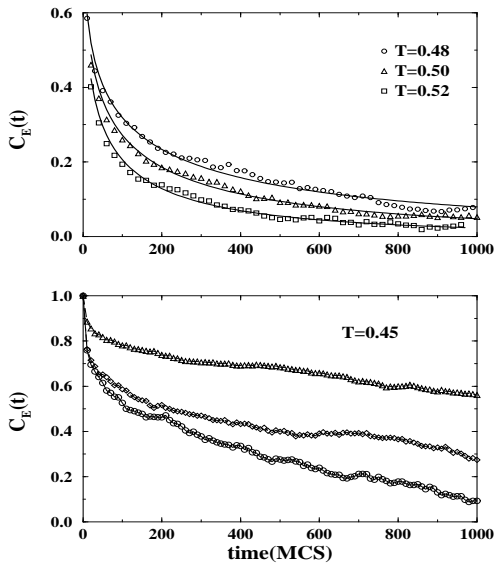}
\vspace{0.1in}
\caption{Top frame shows $C_E (t,t_0)$ for three different temperatures where
$C_E (t,t_0)$ does not depend on $t_0$.  The solid lines are stretched
exponential fits. Bottom frame shows the 
waiting-time dependence of the energy autocorrelation function at $T=0.45$.  The
curves were 
obtained by averaging $C_E (t, t_0 )$ over three different ranges of $t_0$.
From bottom to top, these ranges are $0<t_0 <25000$, $18000<t_0 <48000$ and
$50000<t_0 <80000$.}
\label{waiting}
\end{figure}

As the glass transition is approached, global quantities such as the energy per
spin should exhibit anomalously slow relaxation processes.  Fig (\ref{waiting}) shows the energy autocorrelation 
function $C_{E}(t,t_0)= <E(t_0)E(t+t_0)>$. For quench temperatures between $T=0.6$ and
T=0.47, $C_E (t,t_0)$ is {\it independent} of the time origin $t_0$ and has a stretched exponential form; $\exp^{
-(t/{\tau}_{E})^{\beta}}$.  The stretching-exponent $\beta$ decreases from 0.45 to 0.35 
over this temperature range and the time scale $\tau_{E}$ increases rapidly as is
evident from the top panel of Fig (\ref{waiting}).  
Below $T \simeq 0.47$, the energy autocorrelation function 
depends on the time origin $t_0$, indicating that the equilibration times have
become longer than our observation times.  To illustrate the dependence on the
waiting time $t_0$, we have shown
the autocorrelation function averaged over three different ranges of $t_0$ at
$T=0.45$. The system is seen to relax more slowly for longer waiting times
$t_0$.  This behavior of the energy autocorrelation function is similar to that
of supercooled liquids and the temperature $T=0.47$ is analogous to the
laboratory glass transition temperature at which the equilibration time becomes
longer than the observation time.
In the CTIAFM, the proximity of this
transition to $T^*$ suggests that the glassy dynamics is related to the entropy
vanishing transition.

\begin{figure}[tbhp]
\hspace{0.2in}
\epsfxsize=2.75in \epsfysize=1.8in
\epsfbox{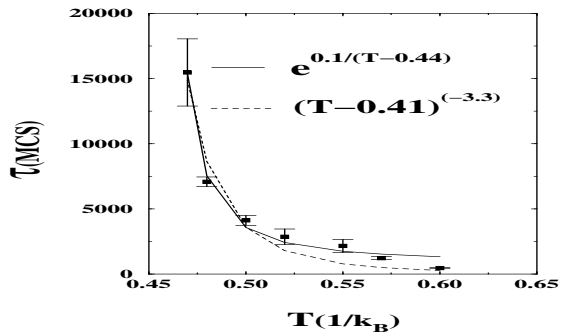}
\caption{Temperature dependence of the string-density  relaxation time.  The
simulation results are shown with error bars.
The solid line is a fit 
to the Vogel-Fulcher form and the dashed line to a power-law.}  
\label{vogel}
\end{figure}

The microscopic picture of the entropy-vanishing transition is 
one where the string density vanishes.  An analysis of the string-density
relaxation can, therefore provide some insight into the nature of the dynamics
at the glass transition. We find that  
the string-density autocorrelation
functions are well described by exponential
relaxations, in contrast to the energy autocorrelation functions.  Fig. (\ref{vogel})
shows the results of our simulations for the relaxation times and fits to a power-law and a Vogel-Fulcher 
form\cite{Vogel}.   Both fits  yield a time-scale divergence at
a temperature  $T \simeq T^*$. 
The $\tau_{E}$ obtained from the stretched exponential fits to $C_E (t,t_0)$
has a temperature dependence which tracks that of the string relaxation time.
This observation  suggests 
that the slow, non-exponential relaxations are a consequence of the freezing of
the string-density relaxation which, in turn, is related to the entropy
vanishing transition.  The static susceptibility associated with the string
density changes only by a factor $\simeq 2$ over the temperature range in which
the time scales change by a factor $\simeq 40$, indicating that the
entropy-vanishing transition suffers from anomalously strong critical slowing down.

\begin{figure}[tbhp]
\epsfxsize=2.5in \epsfysize=3.in
\epsfbox{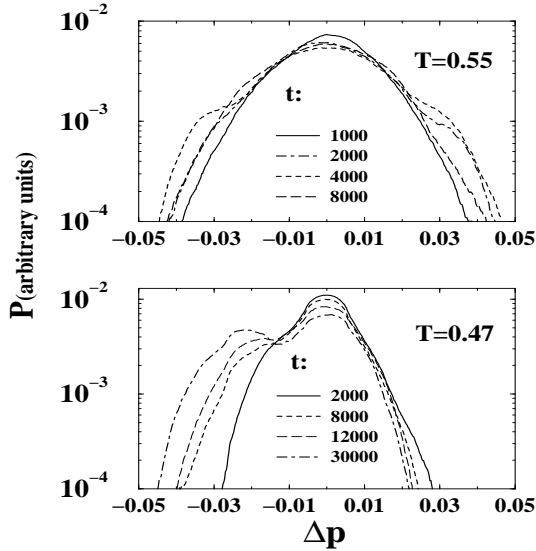}
\caption{Distribution of string-density deviation $\Delta p$ for different
time intervals, $t$, shown at $T=0.55$ and
$T=0.47$. The areas under
the curves have been normalized to unity.}  
\label{non_gauss}
\end{figure}

In order to investigate the nature of the string-relaxations further, we
measured  the distribution, $P(\Delta p)$, of $\Delta p = p(t+t_0 ) - p(t_0)$, the deviation of
the string density in the time interval $t$ (Fig. (\ref{non_gauss})). The distributions were generated by choosing
different time origins $t_0$. The most striking feature of the distributions, observed at temperatures 
close to $T = T^*$, is its non-Gaussian nature at intermediate times.  At
$T=0.55$, the non-Gaussian feature is most pronounced at $t=4000$.  Beyond this time
the distribution relaxes towards a Gaussian and for $t \geq 8000$, the distribution is
time independent. At $T=0.47$, a time-independent behavior is not observed for
times as long as 30,000 and the distributions are non-Gaussian at all intermediate
times. In usual critical-point dynamics\cite{goldenfeld}, one would expect to find a distribution of $\Delta p$
which takes longer to reach its time-independent form as the critical point is
approached and to find non-Gaussian behavior (within the limits of finite-size
cutoffs) in the stationary distribution.  In contrast, we observe the most
pronounced 
non-Gaussian features  at intermediate times.  Drawing an analogy with critical
phenomena, this observation leads us to speculate that there exists 
a time-dependent length scale,  $\xi({\tau})$ 
has a peak at a time $\tau_0 (T)$.  As $T \rightarrow T^*$, the time scale
$\tau_0$ and the height of the peak,  $\xi({\tau}_0)$ appear to diverge.  The
exact nature of the divergence is difficult to extract from the current data.
These apparent divergences indicate that the  thermodynamic transition present in the
zero-defect sector has been replaced by a dynamical transition\cite{Parisi1}.

\paragraph{Connection to real glasses}In conclusion, our study of the CTIAFM provides strong indication that the slow,
glassy
dynamics in this model is associated with an entropy-vanishing transition
involving extended, string-like structures.  These structures are a 
manifestation of the frustration embodied in the nearest-neighbor,
anti-ferromagnetic interactions, and the entropy-vanishing transition is a
consequence of coupling to another degree of freedom, the lattice strain, which
tends to remove the frustration in the system.  The strings are naturally
occurring dynamical heterogeneities since they demarcate regions of fast spin
flips. If the relation between frustration and dynamical heterogeneities is a
generic feature of glass formers, then our observations would suggest that 
the clusters of mobile 
particles observed in Lennard-Jones simulations\cite{Glotzer} and in colloidal
systems\cite{Weitz} should consist of the most frustrated particles in the
system.  In a Lennard-Jones mixture, these should be the particles with the
least number of unlike bonds (if unlike bonds are energetically preferred) and
in a colloidal system, these should be the
particles which have coordinations that are furthest from being icosahedral.  An 
experimental verification of this correlation between geometry and mobility
would be a direct test of the connection between the 
glass-transition and an entropy-vanishing transition involving 
extended structures forced in by frustration.

The work of BC was supported in part by NSF grant number DMR-9815986 and the
work of HY was supported by DOE grant DE-FG02-ER45495.  We would like to thank
R. K. P.  Zia, W. Klein, H. Gould, S. R. Nagel and J. Kondev for many helpful discussions.


\begin{thebibliography}{99}
\bibitem{Glassrev} C. A. Angell, \journal{J. Phys. Chem.}{49}{863}{1988}, M. D. Ediger, C. A. Angell, and S. R. Nagel, \journal{J. Phys. Chem.}{100}{13200}{1996} 
W. G\"{o}tze and L. Sjogren, \journal{Rep. Prog. Phys. B}{55}{241}{1992}
\bibitem{Glotzer} C. Donati, S. C. Glotzer,
P. H. Poole, W. Kob, and S. J. Plimpton,\mpre{60}{3107}{1999} and references therein.
\bibitem{Klein}G. Johnson, A. Mel'cuk, H. Gould, W. Klein and R. Mountain,
\mpre{57}{5707}{1998}.
\bibitem{Ediger}M. T. Cicerone and M. D. Ediger,
\journal{J. Chem. Phys}{103}{5684}{1995} and references therein.
\bibitem{Weitz} E. R. Weeks, J. C. Crocker, A. C. Levitt, A. Schofield, D. A. Weitz, \journal{Science}{287}{627}{2000};W. K. Kegel, A. van Blaaderen, \journal{Science}{287}{290}{2000}.
\bibitem{Adam} G. Adam and J. H. Gibbs, \journal{J. Chem. Phys.}{43}{139}{1965}, 
J. H. Gibbs and E. A. DiMarzio, \journal{J. Chem. Phys.}{28}{373}{1958}.
\bibitem{Mezard}S. Franz and G. Parisi, \mprl{79}{2486}{1997}; M. Mezard and G. Parisi, preprint cond-mat/0002128.
\bibitem{Henley} C. Zeng and C. L. Henley, \mprb{55}{14935}{1997}
\bibitem{wannier}G. H. Wannier, \journal{Phys.~Rev.}{79}{357}{1950}.
\bibitem{houtappel} R. M. F. Houtappel, \journal{Physica}{16}{425}{1950}
\bibitem{Kardar} Z. Y. Chen and M. Kardar, J. Phys. C {\bf 19}, 6825 (1986).
\bibitem{Lei1}Lei Gu, Bulbul Chakraborty, P. L. Garrido, Mohan Phani and
J. L. Lebowitz, Phys. Rev. B {\bf 53}, 11985 (1996); Lei Gu, Ph. D. Thesis, Brandeis
University, 1999.
\bibitem{Dhar} A. Dhar, P. Chaudhuri and C. Dasgupta, \mprb{61}{6227}{2000}.
\bibitem{Zeng} C. Zeng, P. L. Leath and T. Hwa, \mprl{83}{4860}{1999}.
\bibitem{Blote}Henk W. J. Bl\"{o}te and M. Peter Nightingale, \mprb{47}{15046}{1993}.
\bibitem{Kirkpatrick} T. R. Kirkpatrick and P. G. Wolynes,\mpra{35}{3072}{1987}, 
T. R. Kirkpatrick and D. Thirumalai, \mprl{58}{2091}{1987}.
\bibitem{Vogel} H. Vogel, \journal{Phys. Z.}{22}{645}{1921}, G. S. Fulcher,
\journal{J. Am. Ceram. Soc.}{8}{339}{1925}
\bibitem{goldenfeld} N. Goldenfeld, {\it Lectures on Phase Transitions and the Renormalization
Group},(Addison-Wesley, New York, 1992) and Hui Yin, unpublished.
\bibitem{Parisi1}S. Franz, C. Donati,
G. Parisi and S. C. Glotzer, \journal{Phil. Mag. B}{79}{1827}{1999}
\end{thebibliography}
\end{document}